\documentclass[11pt,a4paper]{article}
\usepackage[margin=25mm]{geometry}
\usepackage[T1]{fontenc}
\usepackage{lmodern,microtype,amsmath,amssymb,bm,graphicx,booktabs}
\usepackage[numbers,sort&compress]{natbib}
\usepackage[colorlinks=true,linkcolor=blue,citecolor=blue,urlcolor=blue]{hyperref}
\usepackage{authblk}
\usepackage[font=small,labelfont=bf]{caption}
\usepackage{placeins}
\hypersetup{pdftitle={Nonlinear Dirac Neutrino Oscillations in Matter},pdfauthor={Jianlong Lu, Zhi Hao Quek, Wei Khim Ng, Pulkit Agarwal, Aik Hui Chan, Choo Hiap Oh}}
\newcommand{\id}{\mathbb{I}}

\newcommand{\Hv}{H_{\rm vac}}
\newcommand{\Hm}{H_{\rm mat}}
\newcommand{\Vnc}{V_{\rm NC}}

\newcommand{\La}{\Lambda_{\rm NL}}
\newcommand{\avg}[1]{\langle #1\rangle_\psi}
\newcommand{\Tr}{\operatorname{Tr}}
\title{Nonlinear Dirac Neutrino Oscillations in Matter:\\
Conserved Current, Effective Flavor Dynamics,\\and Trace Sensitivity}
\author[1]{Jianlong Lu\thanks{Corresponding author: \href{mailto:jianlong@nus.edu.sg}{jianlong@nus.edu.sg}.}}
\author[2]{Zhi Hao Quek}
\author[3]{Wei Khim Ng}
\author[2]{Pulkit Agarwal}
\author[2]{Aik Hui Chan}
\author[2]{Choo Hiap Oh}
\affil[1]{\small Department of Mathematics, National University of Singapore, 21 Lower Kent Ridge Road, Singapore 119076}
\affil[2]{\small Department of Physics, National University of Singapore, 21 Lower Kent Ridge Road, Singapore 119077}
\affil[3]{\small Singapore University of Technology and Design, 8 Somapah Road, Singapore 487372}
\date{}
\begin{document}
\maketitle
\begin{abstract}
Matching relativistic dynamics to a consistently normalized flavor amplitude makes neutrino oscillations a quantitative probe of nonlinear propagation. We perform this matching for a flavor-singlet extension of an information-theoretic Dirac action in static matter. The derivative nonlinearity changes the conserved current; a field normalization incorporates this effect and yields, at first order in the nonlinear coupling, $H_{\rm eff}[\psi]=H_0-\eta[\psi]C$, where $C=2H_{\rm vac}+H_{\rm mat}$ and $\eta=2\langle 2\widehat H_{\rm vac}+\widehat V\rangle_\psi/\Lambda_{\rm NL}$. Hats denote the traceful microscopic operators. Consequently, absolute masses and the flavor-universal neutral-current potential can affect relative flavor phases through the nonlinear coefficient. We derive a commutator criterion for its evolution, a conserved functional for constant matter, and a finite two-flavor resonance response. A second scalar-derivative nonlinearity vanishes exactly for a stationary beam with a timelike background. Reproducible two- and three-flavor calculations verify the linear limit, conservation laws, and quadratic convergence of first-order perturbation theory. The resulting benchmark specifies the observable structure and domain of this nonlinear model, enabling correlated comparisons across energy, source flavor, and matter composition with a flavor-coupling structure fixed by the action.
\end{abstract}
\newpage

\section{Introduction}
Neutrino oscillations measure small differences between propagation phases over macroscopic distances. Their interpretation normally assumes linear quantum evolution, with the vacuum mass matrix supplemented by coherent forward scattering in matter~\cite{Wolfenstein,MSW}. A definite propagation model connects a nonlinear relativistic equation to measurable changes in an oscillation experiment and supplies the physical content of a test of linear evolution. In particular, derivative nonlinearities can change the conserved current, and quantities that are unobservable common phases in the linear theory can enter nonlinear couplings.

Nonlinear Dirac equations and their neutrino applications have an established history. Ng and Parwani classified nonpolynomial extensions~\cite{NgDirac}, studied oscillation corrections through nonlinear dispersion relations~\cite{NgProbe}, and constructed information-theoretic actions~\cite{NgInfo}; further dispersion-based constraints were considered in Ref.~\cite{Cheung}. Quek \textit{et al.} subsequently investigated the first action studied here numerically in a low-energy matter setting~\cite{Quek}. Building on these studies, we determine how this particular action reduces to the ultrarelativistic flavor sector when the current, common matter potential, and state dependence are treated together.

Other nonlinear frameworks address different physics. Kaplan and Rajendran constructed a causal state-dependent extension of quantum field theory~\cite{Kaplan}; Gherghetta and Shkerin applied it to neutrino propagation through a Higgs-neutrino interaction~\cite{Gherghetta}. Its interaction through a quantum expectation value differs from the local spinor-bilinear ratios used below. Likewise, neutrino-neutrino forward scattering produces collective mean-field dynamics within standard quantum theory~\cite{Duan}, whereas our model concerns a single coherent beam in prescribed ordinary matter. These distinctions matter when interpreting experimental bounds or statements about causality. In linear propagation, matter-flow invariants organize additive Hamiltonian deformations after removal of the common phase~\cite{LuMatterFlows}. The state-dependent correction derived here instead couples the microscopic trace to a nontrivial flavor generator.

Our contribution is a current-normalized reduction and an explicit analysis of what it predicts. The leading correction is restricted to one flavor direction, $C=2\Hv+\Hm$, with a scalar coefficient determined by a traceful expectation value. This establishes a direct link between a specified relativistic action and three consequences: sensitivity to a common mass-squared shift and neutral-current refraction, a commutator-controlled variation of the nonlinear strength, and a correlated change of vacuum and matter terms. The last effect shifts instantaneous two-flavor maximal mixing while preserving a finite probability response. Numerical examples test the effective equation with its evolving nonlinear coefficient.

The significance is therefore both physical and methodological. A trace removable from a linear oscillation generator can carry physical information in its nonlinear matching coefficient. Local degeneracies with conventional parameter shifts motivate a joint analysis of the correlated dependence across preparations, energies, and compositions; this supplies the information needed to interpret a distortion in an individual probability curve. We derive the action and current in Sec.~\ref{sec:action}, obtain the effective equation in Sec.~\ref{sec:matching}, analyze these consequences in Sec.~\ref{sec:structure}, and present calculations and their interpretation in Secs.~\ref{sec:numerics} and \ref{sec:discussion}.

\section{Action, equations of motion, and conserved current}\label{sec:action}
\subsection{Model and matter conventions}
Let $N=(\nu_1,\ldots,\nu_n)^T$ collect Dirac spinor wave amplitudes in the mass basis, with $M=\operatorname{diag}(m_1,\ldots,m_n)$ and $\nu_f=UN$. We use $n=3$, or $n=2$ for a closed illustrative two-flavor model, and the metric $(+,-,-,-)$. Spinor matrices act trivially in flavor space. In natural units the action is
\begin{align}
\mathcal L={}&\frac{i}{2}\bar N\gamma^\mu\overleftrightarrow{\partial_\mu}N
-\bar N M N-\bar N\gamma^0 V_m P_LN
+\frac{K^2}{4\Lambda_1J}+\frac{S^2}{\Lambda_2J},\label{eq:action}\\
J={}&A_\mu\bar N\gamma^\mu N,\qquad
K=i\bar N\gamma^\mu\overleftrightarrow{\partial_\mu}N,\qquad
S=B^\mu\partial_\mu(\bar N N),\label{eq:bilinears}
\end{align}
where $P_L=(1-\gamma^5)/2$ and $V_m=U^\dagger\widehat VU$. The symmetrized kinetic term differs from $\bar N i\gamma^\mu\partial_\mu N$ by a total derivative. The bilinears in Eq.~\eqref{eq:bilinears} include the sum over flavors \emph{before} the ratios are formed. This is a flavor-singlet extension of the one-spinor actions of Ref.~\cite{NgInfo}; forming separate ratios for individual mass eigenstates specifies a different theory with its own effective generator.

For electrically neutral, unpolarized ordinary matter at rest, the tree-level neutrino potential is
\begin{equation}
\widehat V=V_e P_e+\Vnc\id,\qquad
P_e=\operatorname{diag}(1,0,\ldots,0),\qquad
V_e=\sqrt2G_FN_e,\qquad \Vnc=-\frac{G_F N_n}{\sqrt2}.\label{eq:matter}
\end{equation}
Here $N_e=N_p$ and $N_n$ are number densities. The electron and proton neutral-current vector contributions cancel at this order. We retain $\Vnc$ in the microscopic action to track its nonlinear contribution, while its direct linear contribution is a common active-flavor phase. The benchmark uses the tree-level, unpolarized ordinary-matter potential; polarization, radiative effects, and neutrino self-interactions define further extensions.

We take $A^\mu=(A_0,\bm0)$ with $A_0>0$ and, unless stated otherwise, $B^\mu=(B_0,\bm0)$. Both vectors are constant and dimensionless; $\Lambda_{1,2}$ have dimensions of energy. With the conventional spinor dimension $3/2$, $J$ has dimension three and $K,S$ dimension four, so each nonlinear term in Eq.~\eqref{eq:action} has dimension four. The ratio is defined on configurations with $J\ne0$. For this timelike $A$, $J=A_0N^\dagger N>0$ wherever the amplitude is nonzero. Each action term is homogeneous of degree two under constant amplitude rescaling, and its equation of motion is independent of the arbitrary beam normalization.

We interpret Eq.~\eqref{eq:action} as an effective wave-amplitude model, as in the information-theoretic construction~\cite{NgInfo}. The commuting spinor bilinears specify its wave-amplitude degrees of freedom; embedding this model in a quantized Grassmann-field theory or an electroweak gauge-invariant ultraviolet completion is a separate construction. The second example in Ref.~\cite{NgInfo} used a spacelike $B$; our timelike choice is a distinct restriction whose consequences are stated explicitly below.

\subsection{Variation and normalization}
Define $\chi=K/J$ and $\xi=S/J$. Varying $\bar N$ and $\partial_\mu\bar N$ independently gives
\begin{align}
\mathcal N_1[N]&=i\chi\gamma^\mu\partial_\mu N
+\frac{i}{2}(\partial_\mu\chi)\gamma^\mu N
-\frac{\chi^2}{4}A_\mu\gamma^\mu N,\label{eq:n1}\\
\mathcal N_2[N]&=-2(B\cdot\partial\xi)N-\xi^2 A_\mu\gamma^\mu N,\label{eq:n2}\\
0&=(i\gamma^\mu\partial_\mu-M-\gamma^0V_mP_L)N
+\frac{\mathcal N_1[N]}{\Lambda_1}+\frac{\mathcal N_2[N]}{\Lambda_2}.\label{eq:dirac}
\end{align}
Equation~\eqref{eq:n2} assumes constant $B$. The short variation in Appendix~\ref{app:variation} records the factors and signs. These equations contain higher derivatives through $\chi$ and $\xi$. The propagation analysis selects the perturbative branch connected to the linear Dirac equation, evaluating the nonlinear functionals on that branch at first order. The full higher-derivative initial-value problem is a separate dynamical question.

The global phase symmetry of Eq.~\eqref{eq:action} yields the conserved current
\begin{equation}
\mathcal J^\mu=f\bar N\gamma^\mu N,\qquad
f=1+\frac{\chi}{\Lambda_1},\qquad
\partial_\mu\mathcal J^\mu=0.\label{eq:current}
\end{equation}
The $S$ term adds no phase current because $\bar N N$ is unchanged even by a local phase rotation. In contrast, the $K^2$ term changes the current by $(\chi/\Lambda_1)\bar N\gamma^\mu N$. Hence $\bar N\gamma^\mu N$ alone is generally not conserved. We require $f>0$ on the perturbative branch.

For the first sector, introduce the normalized spinor
\begin{equation}
\mathcal N=\sqrt f\,N.\label{eq:normalization}
\end{equation}
The derivative terms combine exactly as
\begin{equation}
i f\gamma^\mu\partial_\mu N+\frac{i}{2}(\partial_\mu f)\gamma^\mu N
=i\sqrt f\,\gamma^\mu\partial_\mu\mathcal N.\label{eq:cancel}
\end{equation}
Thus, when the second sector is absent, Eq.~\eqref{eq:dirac} is algebraically equivalent to
\begin{equation}
\left[i\gamma^\mu\partial_\mu-\frac{M}{f}
-\gamma^0\frac{V_m}{f}P_L
-\gamma^0\frac{A_0\chi^2}{4\Lambda_1f}\right]\mathcal N=0,\qquad
\mathcal J^\mu=\overline{\mathcal N}\gamma^\mu\mathcal N.\label{eq:canonical}
\end{equation}
The scalar $f$ remains an implicit functional of the field, preserving the nonlinear dynamics. The transformation retains the derivative term at its matching order: if $\chi=O(h)$ and $\partial\chi=O(h^2)$ on an oscillation scale $h$, both $\partial\chi/\Lambda_1$ and the retained nonlinear oscillation correction are of order $h^2/\Lambda_1$.

\section{Ultrarelativistic matching}\label{sec:matching}
\subsection{Stationary beam and leading scalar}
Consider a coherent negative-helicity beam in static matter, with fixed energy $E$, $m_i/E\ll1$, $\|\widehat V\|/E\ll1$, and slowly varying longitudinal envelopes. At zero nonlinear coupling, write
\begin{equation}
N^{(0)}(t,x)=e^{-iEt+iEx}
\begin{pmatrix}\varphi(x)\\r(x)\end{pmatrix}\otimes\zeta_-,\qquad
r=\frac{M}{2E}\varphi+\text{higher relativistic orders}.\label{eq:beam}
\end{equation}
Here $\bm\sigma\cdot\widehat{\bm x}\,\zeta_-=-\zeta_-$. The leading envelope satisfies $i\varphi'=(M^2/2E+V_m)\varphi$. The conserved longitudinal flux of the normalized spinor is the dominant-envelope norm up to relative $O(m^2/E^2)$ corrections. This propagation approximation describes coherent relativistic amplitudes, with flavor-state preparation treated at the corresponding relativistic order~\cite{Giunti}. An exact flavor Fock-space construction and a low-energy flavor-preparation prescription are separate questions. In linear theory, production--propagation--detection analyses identify the conditions for the usual matter evolution~\cite{AkhmedovWilhelm}; propagator treatments recover it for Dirac~\cite{DvornikovDirac} and Majorana~\cite{Dvornikov} neutrinos within their respective approximations.

The linear equation and its adjoint imply the on-shell identity
\begin{equation}
K^{(0)}=2\bar N^{(0)}(M+\gamma^0V_mP_L)N^{(0)}.\label{eq:onshell}
\end{equation}
Using $\bar N^{(0)}MN^{(0)}=\varphi^\dagger M^2\varphi/E$ and $N^{(0)\dagger}N^{(0)}=\varphi^\dagger\varphi$ to leading order gives
\begin{equation}
\chi^{(0)}=\frac{2}{A_0}\,
\frac{\varphi^\dagger(M^2/E+V_m)\varphi}{\varphi^\dagger\varphi}.
\label{eq:chi}
\end{equation}
In the flavor basis define
\begin{equation}
\widehat H_{\rm vac}=U\frac{M^2}{2E}U^\dagger,\qquad
\widehat C=2\widehat H_{\rm vac}+\widehat V,\qquad
q[\psi]=\frac{\psi^\dagger\widehat C\psi}{\psi^\dagger\psi},\qquad
g=\frac{1}{A_0\Lambda_1}=\frac{1}{\La}.\label{eq:definitions}
\end{equation}
Then $\chi^{(0)}=2q/A_0$ and $\eta=\chi^{(0)}/\Lambda_1=2gq$. The overall factor $\sqrt f$ cancels from these leading bilinear ratios. At strict first order, $q$ is evaluated on the linear trajectory of the specified initial state.

\subsection{Effective flavor equation}
Eliminating the small chiral component in Eq.~\eqref{eq:canonical} gives the normalized dominant envelope Hamiltonian
\begin{equation}
\widehat H_{\rm eff}
=\frac{\widehat H_{\rm vac}}{f^2}+\frac{\widehat V}{f}
+\frac{A_0\chi^2}{4\Lambda_1 f}\id
=\widehat H_0-2gq\widehat C+gq^2\id+\cdots,
\qquad \widehat H_0=\widehat H_{\rm vac}+\widehat V.\label{eq:htraceful}
\end{equation}
All displayed nonlinear terms are first order in $g$; the last term is a common vector potential. The two factors of $f^{-1}$ in the vacuum part arise from the two mass insertions, while the matter part has one. This is the origin of the relative coefficients in $\widehat C$.

For every $n\times n$ matrix $X$, let $X_{\rm tl}=X-(\Tr X/n)\id$. We consistently use
\begin{equation}
\Hv=(\widehat H_{\rm vac})_{\rm tl},\quad
\Hm=(\widehat V)_{\rm tl}=V_e(P_e-\id/n),\quad
H_0=\Hv+\Hm,\quad C=2\Hv+\Hm.\label{eq:traceless}
\end{equation}
Removing the scalar phase from Eq.~\eqref{eq:htraceful} yields
\begin{equation}
i\frac{d\psi}{dx}=\bigl(H_0-\eta[\psi]C\bigr)\psi,
\qquad \eta[\psi]=2g\bigl(h_{\rm tr}+\avg{C}\bigr),\qquad
h_{\rm tr}=\frac{\sum_i m_i^2}{nE}+\frac{V_e}{n}+\Vnc.
\label{eq:main}
\end{equation}
Equation~\eqref{eq:main} is the leading effective equation studied below. Even with a zero lightest mass, the vacuum trace generally remains finite. Combining the traceless vacuum matrix with $h_{\rm tr}$ counts each contribution once. Computing $q$ directly from $\widehat C$ implements the same prescription.

The microscopic trace is fixed before the phase removal. The rephasing of the final envelope leaves $q$ invariant, so scalar terms in Eq.~\eqref{eq:htraceful} can be removed even when state-dependent. The expectation value therefore retains $\widehat C$. A consistent local rephasing at the action level also transforms the derivative bilinear $K$, preserving the same microscopic model. The need to distinguish physical phase effects from representation artifacts also arises in linear neutrino geometric-phase calculations~\cite{LuPhase}.

The expansion is organized by $m/E$, $\|\widehat V\|/E$, and $|g|h$, with $h$ a bound on the traceful mass and matter scales. Leading linear terms are $O(h)$ and the displayed nonlinear terms are $O(gh^2)$. Away from cancellations, ordinary relativistic corrections are suppressed by $m^2/E^2$ or $\|\widehat V\|/E$, and the omitted nonlinear feedback terms by an additional $|g|h$. These are separate expansions. A numerical solution of Eq.~\eqref{eq:main} resums feedback within the leading effective equation. Full accuracy at $O(g^2)$ additionally requires matching the Dirac theory at that order.

\subsection{Second nonlinearity and the stationary restriction}
For the beam of Eq.~\eqref{eq:beam}, all bilinears are independent of $t$ even when they vary with $x$. Therefore
\begin{equation}
B^\mu=(B_0,\bm0),\quad N(t,x)=e^{-iEt}n(x)
\quad\Longrightarrow\quad S=B_0\partial_t(\bar N N)=0,\quad \xi=0,\quad\mathcal N_2=0.
\label{eq:secondzero}
\end{equation}
This identity holds exactly throughout the stationary sector. For a slowly varying time-dependent packet, or a background with a spatial component, $\bar NN/(N^\dagger N)=O(m/E)$ instead implies, schematically,
\begin{equation}
\xi=O\!\left(\frac{|B|}{A_0}\frac{m}{E}\omega_{\rm env}\right),\qquad
\delta H_2=O\!\left(\frac{|B|^2}{A_0\Lambda_2}
\frac{m^2}{E^2}\omega_{\rm env}^2\right).\label{eq:secondscale}
\end{equation}
The second estimate includes projection of the scalar term in Eq.~\eqref{eq:n2}, which supplies another factor $m/E$; its vector term has the same scaling. It assumes derivatives of the envelope and its logarithmic density remain on the scale $\omega_{\rm env}$. Arbitrary packets and backgrounds require the corresponding time-dependent or directional analysis.

\section{Observable structure and analytic consequences}\label{sec:structure}
\subsection{Trace sensitivity and its physical meaning}
The scalar trace in Eq.~\eqref{eq:main} has a measurable role only through multiplication by the nontrivial flavor generator $C$. For a common change of the physical masses squared and a common matter-potential change, at fixed $U$, $E$, and splittings,
\begin{align}
m_i^2&\longrightarrow m_i^2+\delta m_0^2:
&\delta\eta&=\frac{2g\delta m_0^2}{E},
&\delta H_{\rm eff}&=-\frac{2g\delta m_0^2}{E}C,\label{eq:massshift}\\
\Vnc&\longrightarrow\Vnc+\delta V_{\rm NC}:
&\delta\eta&=2g\delta V_{\rm NC},
&\delta H_{\rm eff}&=-2g\delta V_{\rm NC}C.\label{eq:ncshift}
\end{align}
These formulas compare physically distinct microscopic inputs at a fixed phase convention. At first order the same linear trajectory can be used on both sides. In particular, two media with the same electron density but different neutron densities have the same standard active-flavor Hamiltonian and can have different nonlinear corrections in this model. This model-specific composition dependence supplies a discriminant whose experimental reach can be quantified with a dedicated sensitivity analysis.

An unknown absolute mass scale is consequently a physical nuisance parameter in a fit to $g$.'' The observable combination can be small through cancellations among the vacuum, charged-current, and neutral-current terms. The sign of $q$ depends on the state and medium even for positive masses and positive $\Lambda_{1}$, allowing both increases and decreases in oscillation probabilities.

\subsection{State dependence, conserved functional, and vacuum limit}
The Hermiticity of $H_0$ and $C$, and reality of $\eta$, give $d(\psi^\dagger\psi)/dx=0$. For constant matter, the effective equation also conserves
\begin{equation}
\mathcal E[\psi]=\avg{H_0}-g\bigl(h_{\rm tr}+\avg{C}\bigr)^2.
\label{eq:energy}
\end{equation}
Indeed, $q'=i\avg{[H_0,C]}$ and $\avg{H_0}'=2gq\,q'$, so $\mathcal E'=0$. The identity term in Eq.~\eqref{eq:htraceful} is also obtained by varying the homogeneous functional $\psi^\dagger\widehat H_0\psi-g(\psi^\dagger\widehat C\psi)^2/(\psi^\dagger\psi)$. Thus Eq.~\eqref{eq:energy} provides a second numerical diagnostic beyond norm conservation.

More explicitly,
\begin{equation}
\eta'=-2ig\avg{[\Hv,\Hm]}\qquad\text{(constant matter)}.\label{eq:commutator}
\end{equation}
Hence $[\Hv,\Hm]=0$ is sufficient for a constant coefficient for every preparation. For nonzero $g$, it is also necessary if constancy is required for every initial pure state: a Hermitian operator whose expectation vanishes in every state is zero. Particular trajectories may still have constant $\eta$ when the commutator is nonzero. At constant density, noncommuting vacuum and matter terms can sustain an evolving coefficient. In varying matter the additional explicit term gives
\begin{equation}
\eta'=2g\left(\avg{\widehat C'}-i\avg{[\Hv,\Hm]}\right).
\end{equation}

In vacuum, $\eta$ is constant for each prepared state, and Eq.~\eqref{eq:main} is solved exactly within the leading effective model by
\begin{equation}
\psi(x)=e^{-i(1-2\eta_\alpha)\Hv x}e_\alpha,\qquad
\eta_\alpha=\frac{2g}{E}\sum_i|U_{\alpha i}|^2m_i^2.
\label{eq:vacuum}
\end{equation}
The nonlinear frequency correction scales as $E^{-2}$ at fixed masses and $g$. For one preparation at one energy it is degenerate with a common rescaling of the measured mass-squared splittings. Across energies or source flavors the rescaling is correlated and generally changes. The restricted matrix structure therefore supplies correlated predictions whose distinguishability from linear alternatives depends on the available preparations and spectral coverage.

\subsection{Preparation-specific perturbation theory}
For an initial flavor $e_\alpha$, let $\psi_\alpha^{(0)}(x)=S_0(x)e_\alpha$, where $S_0(x)=e^{-iH_0x}$ for constant matter. Define $\eta_\alpha^{(0)}(x)=2g\langle\widehat C\rangle_{\psi_\alpha^{(0)}(x)}$. Then
\begin{align}
\delta\psi_\alpha(L)&=i\int_0^L S_0(L-x)\eta_\alpha^{(0)}(x)C\psi_\alpha^{(0)}(x)\,dx,\label{eq:duhamel}\\
P_{\alpha\beta}^{(1)}&=|\psi_{\alpha,\beta}^{(0)}(L)|^2
+2\operatorname{Re}\left[\psi_{\alpha,\beta}^{(0)*}(L)\delta\psi_{\alpha,\beta}(L)\right].\label{eq:probpert}
\end{align}
The self-consistent evolution preserves the norm. Near a probability extremum, the truncated first-order expression can lie slightly outside $[0,1]$ by higher-order terms; its accuracy is assessed against the self-consistent result.
Each initial state supplies its own scalar history and is propagated individually. The resulting nonlinear map preserves the norm of each trajectory; overlaps between distinct trajectories can evolve, so its action on superpositions requires evolving their full initial states. For varying density, $S_0(L-x)$ is replaced by the two-point linear propagator $S_0(L,x)$.

Local matching requires the relativistic hierarchies and $|g|h\ll1$, with $h$ bounding the traceful mass and matter scales as in Sec.~\ref{sec:matching}. The trajectory diagnostic $|\eta|\ll1$ complements this condition, particularly when contributions to $q$ cancel. Over a finite baseline, the magnitude of the first-order response is measured by
\begin{equation}
\epsilon_L=\int_0^L|\eta_\alpha^{(0)}(x)|\,\|C(x)\|\,dx\ll1.
\label{eq:secular}
\end{equation}
At first order, $\|\delta\psi\|\le\epsilon_L$ and $|\delta P_{\alpha\beta}|\le2\epsilon_L$. For constant matter, $\epsilon_L\le S_L\equiv2|g|L\|\widehat C\|\|C\|$. Requiring $S_L\ll1$ supplies a uniform sufficient regime for the self-consistent expansion, including nonlinear feedback. Indeed, normalized states satisfy $|q[\psi]-q[\phi]|\le2\|C\|\|\psi-\phi\|$, and a Duhamel--Gronwall estimate gives $\|\psi(L)-\psi^{(0)}(L)\|\le\epsilon_L\exp(4|g|\|C\|^2L)$. Since $\|C\|\le2\|\widehat C\|$, the exponent is at most $4S_L$. The trajectory-specific $\epsilon_L$ bounds the first-order term; the uniform condition also controls amplification near cancellations. These accumulated quantities complement the local ratio $\|\eta C\|/\|H_0\|$, because a locally small correction can build up a substantial phase over a long baseline.

Replacing $\eta(x)$ by $\eta(0)$ incurs a separate error. Equation~\eqref{eq:commutator} implies
\begin{equation}
|\eta(x)-\eta(0)|\le2|g|x\|[\Hv,\Hm]\|.
\end{equation}
The corresponding first-order amplitude error is at most $|g|L^2\|C\|\|[\Hv,\Hm]\|$. This bound quantifies the accuracy of freezing the coefficient and identifies its controlled regime at a given density and baseline.

\subsection{Two flavors and finite resonance response}
For two flavors let $\Delta=\Delta m^2/(2E)$, $a=\Delta\sin2\theta$, and $d=\Delta\cos2\theta-V_e$. Then
\begin{equation}
H_0=\frac12(a\sigma_1-d\sigma_3),\qquad
C=a\sigma_1+\left(-\Delta\cos2\theta+\frac{V_e}{2}\right)\sigma_3.
\end{equation}
If the coefficient is frozen, the Pauli components and probability are
\begin{align}
B_x&=a(1-2\eta),\qquad
B_z=-\Delta\cos2\theta(1-2\eta)+V_e(1-\eta),\label{eq:bvec}\\
\Omega&=\sqrt{B_x^2+B_z^2},\qquad
P_{ee}=1-\frac{B_x^2}{\Omega^2}\sin^2\frac{\Omega L}{2}.
\label{eq:twoexact}
\end{align}
The formula is exact for this frozen generator. For a varying coefficient the same components define an instantaneous generator; the full probability follows from propagation along its evolving trajectory.

Instantaneous maximal mixing, $B_z=0$, obeys
\begin{equation}
V_e=\Delta\cos2\theta\frac{1-2\eta}{1-\eta}
=\Delta\cos2\theta(1-\eta)+O(\eta^2).\label{eq:resonance}
\end{equation}
This is implicit because $\eta$ depends on the state, energy, and density. Locating a baseline-dependent probability maximum or establishing adiabatic conversion additionally uses the full trajectory.

The apparent pole in $\tan2\theta_m$ is a coordinate singularity. At fixed $E,V_e$ the first-order frozen-coefficient changes are
\begin{align}
\delta\theta_m&=\frac{\eta aV_e}{2(a^2+d^2)},\label{eq:mixshift}\\
\delta\Omega&=-\eta\frac{2\Delta^2-3\Delta V_e\cos2\theta+V_e^2}{\sqrt{a^2+d^2}}.\label{eq:freqshift}
\end{align}
For nonzero $a$, both are finite at the linear resonance $d=0$. At that point $\delta\Omega=-2\eta|a|$, while the mixing amplitude $\sin^22\theta_m$ has no first-order shift because it is already at its maximum. The remaining leading change in the frozen probability comes from the phase; for $a>0$, $\delta P_{ee}=\eta aL\sin(aL)$. Narrow mixing resonances can be sensitive to small shifts. The resulting $|\Delta P|$ depends on the oscillation phase and baseline, while its statistical significance also depends on the measured event spectrum. The bound in Eq.~\eqref{eq:secular} remains finite through resonance.

For a self-consistent alternative, set $\bm r=\psi^\dagger\bm\sigma\psi$ for a unit-normalized two-component state and write $C=\bm c\cdot\bm\sigma/2$. Then
\begin{equation}
\bm r'=\left[\bm b_0-2g\left(h_{\rm tr}+\frac{\bm c\cdot\bm r}{2}\right)\bm c\right]\times\bm r,
\qquad \bm b_0=(a,0,-d).\label{eq:bloch}
\end{equation}
This equation preserves $|\bm r|$ and $\mathcal E$, and offers an independent implementation of the nonlinear two-flavor dynamics.

\section{Reproducible constant-density calculations}\label{sec:numerics}
\subsection{Inputs and numerical methods}
We use two fully specified neutrino-propagation benchmarks to expose the analytic structure and quantify its response before an experiment-specific fit. The two-flavor example has $\Delta m^2=7.5\times10^{-5}\,{\rm eV}^2$, $\theta=33.4^\circ$, $m_1=0$, $\rho=100\,{\rm g\,cm^{-3}}$, $Y_e=0.5$, $L=200\,{\rm km}$, and $E=2$--$15\,{\rm MeV}$. Its homogeneous density isolates the response at a solar-relevant matter scale; a solar prediction additionally uses the density profile. The three-flavor example uses normal ordering with
\begin{equation}
\begin{gathered}
\Delta m_{21}^2=7.5\times10^{-5}\,{\rm eV}^2,\qquad
\Delta m_{31}^2=2.5\times10^{-3}\,{\rm eV}^2,\\
(\theta_{12},\theta_{13},\theta_{23},\delta_{\rm CP})=(33.4^\circ,8.6^\circ,49.0^\circ,195^\circ),
\end{gathered}
\end{equation}
$m_1=0$, $\rho=2.8\,{\rm g\,cm^{-3}}$, $Y_e=0.5$, $L=1300\,{\rm km}$, and $E=0.5$--$8\,{\rm GeV}$. These fixed reference values provide representative oscillation inputs independently of a particular global-fit point; contemporary parameter correlations are discussed in Ref.~\cite{NuFit}. We use the conventional PMNS parametrization $U=R_{23}U_{13}(\delta_{\rm CP})R_{12}$.

For both examples, $N_e=\rho Y_e/m_u$ and $N_n=\rho(1-Y_e)/m_u$, with $m_u$ the atomic mass unit. Numerically,
\begin{equation}
V_e=7.63247\times10^{-14}\,{\rm eV}\,
\frac{\rho}{\mathrm{g\,cm^{-3}}}Y_e,\qquad
\Vnc=-\frac{1-Y_e}{2Y_e}V_e,\qquad
1\,{\rm km}=5.067730716\times10^9\,{\rm eV}^{-1}.
\end{equation}
The illustrative coupling is $\La=10^{-8}\,{\rm eV}$. This low effective scale makes the response of the specified wave-amplitude model visible in the benchmarks. Its interpretation as an experimental bound or a heavy-particle threshold requires the corresponding likelihood or microscopic completion. The relevant expansion parameter is $|g|h$, with additional checks below. Interpreting $\La$ as an ultraviolet cutoff would require a completion and a separate validity analysis.

The Python implementation integrates Eq.~\eqref{eq:main} with DOP853, relative tolerance $2\times10^{-11}$ and absolute tolerance $2\times10^{-13}$. It simultaneously evolves the linear trajectory and the inhomogeneous first-order correction, evaluates the linear result independently by matrix exponentiation, and samples each trajectory at 401 positions for conservation diagnostics. The energy grids contain 161 and 151 points, respectively. These are probability-level calculations; detector smearing, rate normalization, and a statistical likelihood enter the event-level analysis described in Sec.~\ref{sec:discussion}.

All microscopic matrices are constructed in eV; multiplication by $5.067730716\times10^9$ converts the generator for integration in km. The code builds $\widehat C$ directly from the physical masses and the full matter potential, and subtracts traces only for $H_0$ and $C$.

\subsection{Two-flavor response and the frozen-coefficient approximation}
\begin{figure}[!htbp]
\centering\includegraphics[width=.79\textwidth]{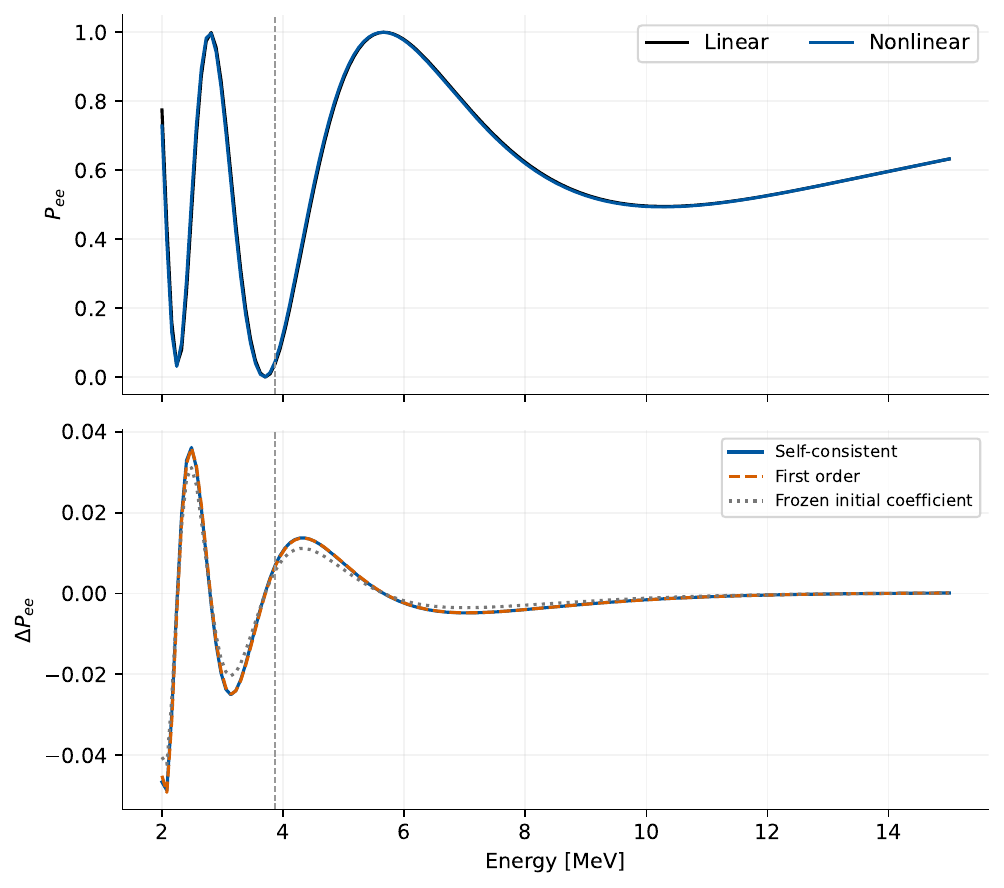}
\caption{Two-flavor benchmark with $L=200\,{\rm km}$, $\rho=100\,{\rm g\,cm^{-3}}$, $Y_e=0.5$, and $\La=10^{-8}\,{\rm eV}$. The upper panel shows survival probabilities. The lower panel shows the absolute distortion relative to the linear result for self-consistent evolution, the first-order expansion, and the frozen initial coefficient. The dashed vertical line denotes the linear mixing resonance; the probability extrema also depend on the baseline. Both charged-current and neutral-current potentials enter $q$.}\label{fig:two}
\end{figure}

Figure~\ref{fig:two} compares the linear and self-consistent $P_{ee}$, then resolves their difference against first-order perturbation theory and a coefficient frozen at its initial value. The linear resonance is at $E=3.871\,{\rm MeV}$, marked by a vertical line. The oscillatory sign changes and energy dependence display the joint roles of the resonance shift and accumulated phase in the probability distortion. The self-consistent curve gives the numerical nonlinear prediction, while the frozen curve quantifies the accuracy of holding the coefficient at its initial value.

\subsection{Three flavors and trace dependence}
\begin{figure}[!htbp]
\centering\includegraphics[width=\textwidth]{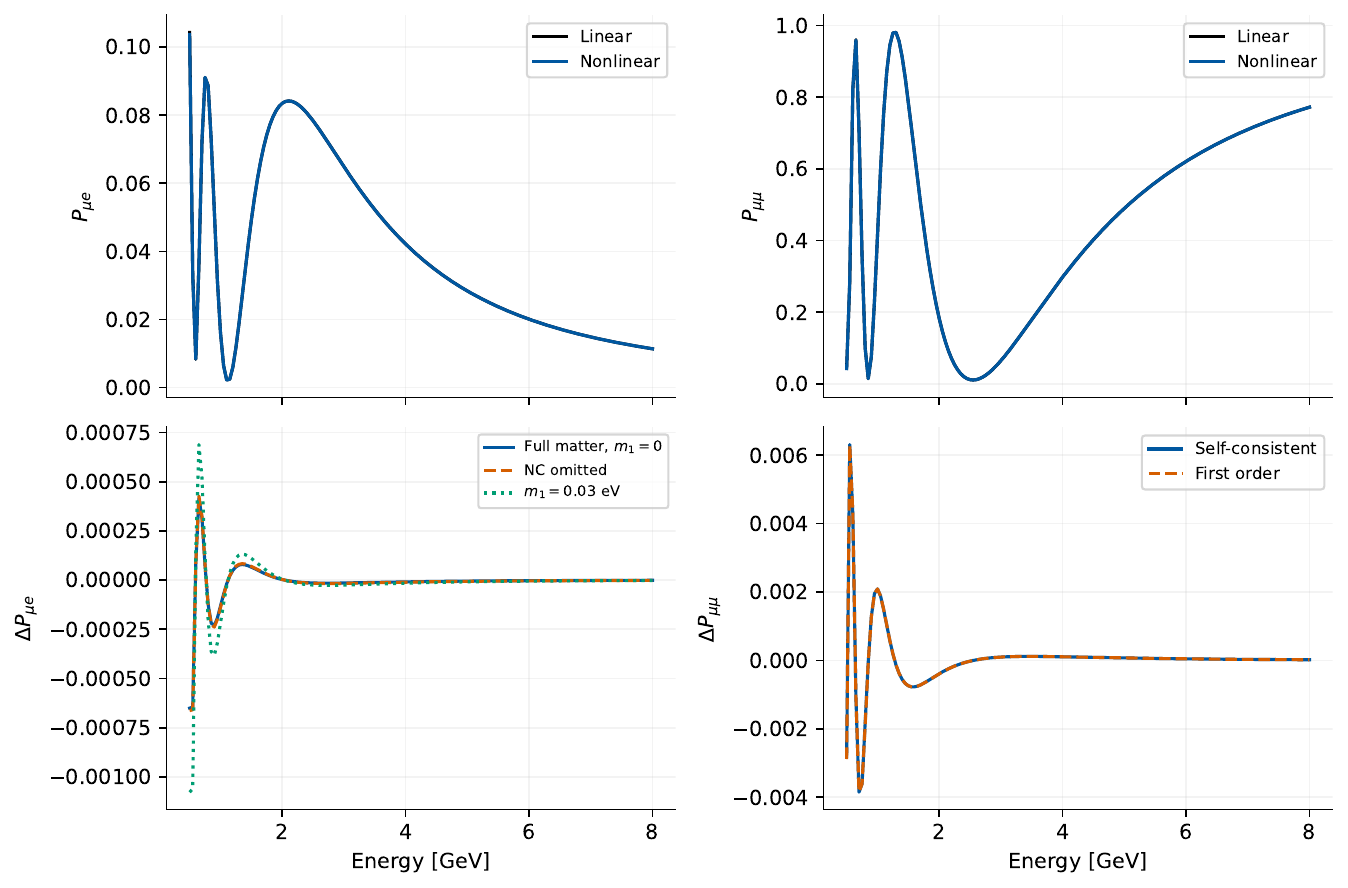}
\caption{Three-flavor constant-density benchmark at $L=1300\,{\rm km}$, with the parameters specified in the text and $\La=10^{-8}\,{\rm eV}$. Upper panels: $P_{\mu e}$ and $P_{\mu\mu}$. Lower left: appearance distortions with full matter and $m_1=0$, with the neutral-current term omitted, and with $m_1=0.03\,{\rm eV}$ and full matter. Lower right: self-consistent and first-order disappearance distortions. Differences are absolute probabilities; experimental significance is obtained from an event-level likelihood.}\label{fig:three}
\end{figure}

Figure~\ref{fig:three} displays $\nu_\mu$ appearance and disappearance. Its lower left panel compares the physical full-matter calculation with two diagnostic settings: omission of $\Vnc$ from $q$, and a common mass-squared shift corresponding to $m_1=0.03\,{\rm eV}$ at fixed splittings. Omitting $\Vnc$ changes the microscopic model and isolates its physical contribution to the nonlinear coefficient. All three cases have the same linear probabilities, but their nonlinear differences resolve the trace dependence in Eqs.~\eqref{eq:massshift} and \eqref{eq:ncshift}. The appearance and disappearance channels respond differently despite sharing the same scalar coupling and fixed generator.

\subsection{Validation and perturbative control}
\begin{figure}[!htbp]
\centering\includegraphics[width=.97\textwidth]{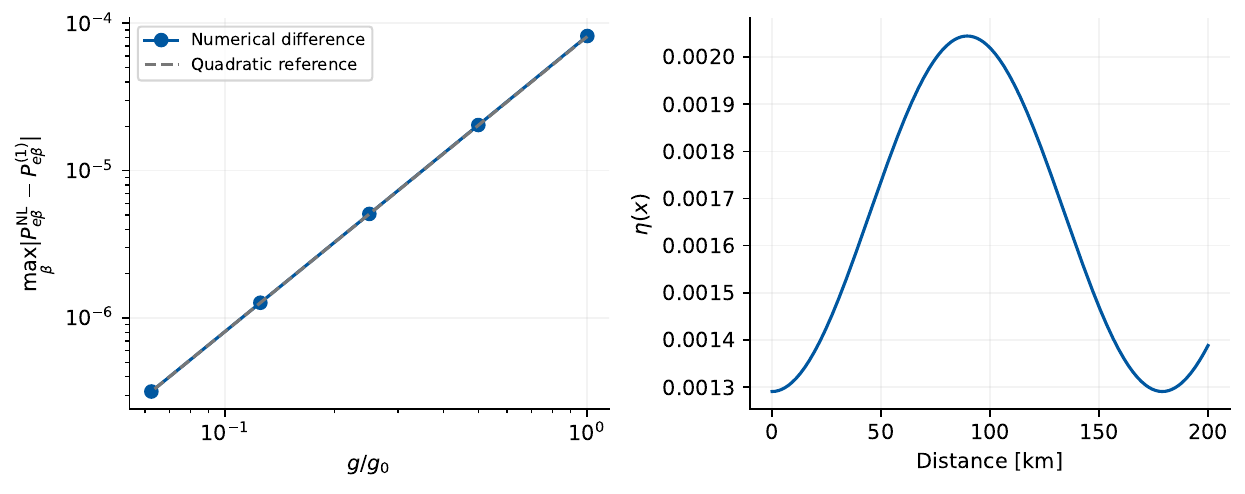}
\caption{Left: maximum channel difference between the self-consistent probability and its first-order expansion at $E=5\,{\rm MeV}$, with $g_0=(10^{-8}\,{\rm eV})^{-1}$. The quadratic reference has been normalized to the smallest-coupling point. Right: state-dependent coefficient along the same two-flavor benchmark at $g=g_0$. These checks quantify the dynamics of the leading effective equation.}\label{fig:validation}
\end{figure}

The implementation checks the analytic two-flavor linear formula, the exact effective-model vacuum solution~\eqref{eq:vacuum}, and an independent integration of Eq.~\eqref{eq:bloch}. At fixed $E=5\,{\rm MeV}$ and $L=200\,{\rm km}$, the coupling is successively halved to test the order of the difference between self-consistent and first-order probabilities. Figure~\ref{fig:validation} also shows $\eta(x)$ at this energy: its variation directly demonstrates the evolving nonlinear coefficient at constant density.
\FloatBarrier
Table~\ref{tab:checks} gives the maxima on the stated energy grids for the default full-matter, zero-lightest-mass benchmarks. The largest absolute two-flavor distortion is $0.0488$; the three-flavor appearance and disappearance maxima are $6.51\times10^{-4}$ and $6.31\times10^{-3}$, respectively. These maxima quantify the response on the stated grids; experimental sensitivity follows from an event-level analysis. The two-flavor sufficient accumulated bound reaches $0.135$, and the corresponding first-order discrepancy is below $0.002$ in absolute probability. The discrepancy is consistent with finite-baseline corrections and is quantified alongside the local smallness ratio.

\begin{table}[!htbp]
\centering
\caption{Numerical diagnostics for the default benchmarks. The energy drift is divided by $\|H_0\|$ at each energy. The accumulated bound is the analytical sufficient bound following Eq.~\eqref{eq:secular}; the last row gives the measured first-order discrepancy.}
\label{tab:checks}
\begin{tabular}{lrr}\toprule
Diagnostic & Two flavors & Three flavors\\\midrule
Maximum $|\eta|$ & $3.39\times10^{-3}$ & $5.57\times10^{-4}$\\
Maximum $\|\eta C\|/\|H_0\|$ & $6.97\times10^{-3}$ & $1.13\times10^{-3}$\\
Accumulated bound & $1.35\times10^{-1}$ & $2.12\times10^{-2}$\\
Maximum $|\psi^\dagger\psi-1|$ & $8.30\times10^{-11}$ & $5.38\times10^{-11}$\\
Scaled $\mathcal E$ drift & $4.32\times10^{-13}$ & $5.92\times10^{-11}$\\
Maximum $|P^{\rm NL}-P^{(1)}|$ & $1.95\times10^{-3}$ & $7.09\times10^{-5}$\\
\bottomrule\end{tabular}
\end{table}

The linear two-flavor formula agrees with matrix exponentiation to $2.22\times10^{-15}$ in absolute probability. The vacuum formula and independent Bloch solver agree with the complex-amplitude integration to $4.67\times10^{-12}$ and $1.72\times10^{-12}$, respectively. A log--log fit of the coupling-halving errors gives a slope of $2.003$, consistent with the expected $O(g^2)$ difference. The common-mass trace-shift identity is verified to $5.21\times10^{-26}\,{\rm eV}$ in the tested matrices.

In addition to the plotted results, it verifies numerically that a common shift $m_i^2\to m_i^2+\delta m_0^2$ changes $\widehat C$ by $(\delta m_0^2/E)\id$. The vacuum test, independent Bloch evolution, linear analytic comparison, and coupling-halving study test complementary aspects of the effective dynamics. The current-normalization argument, Euler--Lagrange equations, and stationary-sector identity follow analytically. Well-posedness, stability, and ultraviolet consistency of the full higher-derivative theory require a corresponding analysis of that theory.

\section{Phenomenological interpretation and scope}\label{sec:discussion}
Matching fixes the flavor direction of the correction and its off-diagonal structure. This makes Eq.~\eqref{eq:main} a restricted benchmark against which to compare more general propagation effects, including nonstandard interactions~\cite{Ohlsson}. For a frozen coefficient it is algebraically identical to replacing $\Hv\to(1-2\eta)\Hv$ and $\Hm\to(1-\eta)\Hm$. Over a restricted energy interval and one source preparation, standard-parameter and density shifts can therefore mimic part of the effect. The distinguishing information is the common functional relation $\eta=2g\langle\widehat C\rangle$, tested across channels, energies, source flavors, and medium compositions. A fitted signal would require marginalization over the standard mixing parameters, the physical absolute mass, and matter uncertainties.

\subsection{Correlated spectral, source, and composition tests}
Equation~\eqref{eq:vacuum} makes the vacuum target more specific than a generic shift of oscillation parameters. Define $m_\alpha^2=\sum_i|U_{\alpha i}|^2m_i^2$. For a fixed initial flavor and energy, the apparent splittings inferred from the propagation phases satisfy, within the leading effective model,
\begin{equation*}
(\Delta m_{ij}^2)_{\mathrm{app},\alpha}(E)
=\left(1-\frac{4g m_\alpha^2}{E}\right)\Delta m_{ij}^2.
\end{equation*}
All splittings acquire the same fractional shift at that setting, so their ratios remain unchanged. An energy-resolved test instead looks for the correlated $1/E$ drift of the apparent splittings, with energy calibration treated as a nuisance. Comparing two source flavors at the same energy gives
\begin{equation*}
\eta_\alpha-\eta_\gamma
=\frac{2g}{E}\sum_i\bigl(|U_{\alpha i}|^2-|U_{\gamma i}|^2\bigr)\Delta m_{i1}^2,
\end{equation*}
where the common absolute-mass contribution cancels by row normalization of $U$. This provides a mass-offset-independent consistency relation for controlled source preparations and propagation conditions. Homogeneity of the action makes these single-particle probabilities independent of the overall beam intensity.

For electron-flavor preparation, $m_e^2$ is the same mass combination denoted by $m_\beta^2$ in standard kinematic beta-decay analyses~\cite{KATRIN259}. This gives a possible connection between propagation and absolute-mass measurements. A joint interpretation with KATRIN requires a consistent electron-antineutrino sector, including the transformation of the external backgrounds under charge conjugation, and production and endpoint kinematics compatible with the endpoint analysis. These ingredients complement the ultrarelativistic propagation matching. A propagation fit can expose this connection by displaying its dependence on the lightest mass. At the linear null hypothesis $g=0$, oscillation data determine mass-squared differences and leave the common mass unconstrained.

The composition test can likewise be expressed directly in quantities along a neutrino trajectory. With $p_e(x)=\langle P_e\rangle_\psi$ and $M_f^2=UM^2U^\dagger$, Eq.~\eqref{eq:definitions} gives
\begin{equation*}
q(x)=\frac{\langle M_f^2\rangle_\psi}{E}
+V_e(x)\left[p_e(x)-\frac{1-Y_e(x)}{2Y_e(x)}\right].
\end{equation*}
At $Y_e=1/2$, the matter contribution is $V_e(p_e-1/2)$: it can change sign during flavor conversion and can cancel the vacuum contribution. Both electron and neutron densities therefore enter the nonlinear coefficient. For a physical comparison at fixed electron density, variations of neutron density must be separated from uncertainties in the density profile. The paired rescalings of vacuum and matter terms described above also distinguish this benchmark from an independently fitted, state-independent NSI matrix; a constraint on $g$ follows from fitting this correlated propagation model to the data.

\subsection{Experimental settings and event spectra}
The three-flavor benchmark uses the baseline and energy scales relevant to DUNE, whose broad energy spectrum and joint electron-appearance and muon-disappearance samples make spectral consistency tests natural~\cite{DUNEPhysics}. It identifies correlated channel distortions that can be carried into a DUNE event-level likelihood. A fit would allow $\delta_{\rm CP}$, $\theta_{23}$, the atmospheric splitting, the mass ordering, and matter uncertainties to vary, since a nonlinear distortion can otherwise be absorbed into their fitted values. Antineutrino components, including wrong-sign beam backgrounds, require the charge-conjugated extension discussed below before a complete accelerator analysis is possible.

The MeV benchmark connects to the vacuum-to-matter transition probed by solar-neutrino measurements~\cite{BorexinoPP}. A solar prediction additionally requires production-region averaging, a varying solar composition, and a justified treatment of wave-packet separation and subsequent Earth propagation. For nonlinear pure-state propagation, these steps require a specified preparation ensemble and a treatment of coherence throughout the trajectory. Reactor spectral measurements, including JUNO~\cite{JUNOFirst}, offer a complementary low-energy setting once the antineutrino equation is established. Their published linear-theory parameter uncertainties motivate a dedicated fit of the charge-conjugated model.

For independently prepared coherent neutrinos, assume that production and detection amplitudes retain their standard forms to the accuracy of the propagation analysis. With this prescription, an explicit bridge to a reconstructed event bin $b$ is
\begin{equation*}
\mu_b=\mathcal X\sum_{\alpha,\beta}\int dE\,
\Phi_\alpha(E)\,\sigma_\beta(E)\,R_{b\beta}(E)\,
P_{\alpha\beta}(E,L;g,\vartheta,\zeta)+B_b.
\end{equation*}
Here $\mathcal X$ contains exposure and target normalization, $\Phi_\alpha$ is the source spectrum, $\sigma_\beta$ the detection cross section, $R_{b\beta}$ the response including efficiency, and $B_b$ the background. The parameters $\vartheta$ contain mixing and masses, while $\zeta$ collects matter and experimental nuisances; fluxes, cross sections, and response functions can also depend on $\zeta$. Each prepared component is evolved separately before probabilities are averaged. This prescription specifies a classical ensemble of preparations; a general nonlinear mixed-state law requires an additional dynamical definition. A constrained count likelihood can then test whether the response $\partial\mu_b/\partial g$ is distinguishable from parameter and systematic shifts across bins~\cite{LuEstimation}. The probability curves provide the propagation input for such an event fit and its eventual exclusion limits.

\subsection{Sensitivity scale and physical scope}

The scale dependence identifies the phenomenological regime of this operator. For $q=O(10^{-12}\,{\rm eV})$ and $\La=1\,{\rm GeV}$, $|\eta|=O(10^{-21})$, placing a high effective scale in a strongly suppressed terrestrial regime. In vacuum its frequency correction decreases as $E^{-2}$, unlike higher-energy growth possible for other nonlinear Dirac structures~\cite{NgProbe}. The figures isolate mathematical response at a chosen low effective scale. A realistic sensitivity study must first establish the domain of the wave-amplitude model, then use finite-resolution event spectra and nuisance correlations; absolute event distortions and their uncertainties determine sensitivity even near probability zeros~\cite{LuEstimation}.

There are two distinct layers of consistency. At the level studied here, the action supplies a conserved current, the normalized effective equation preserves the norm, and its perturbative branch can be checked quantitatively. A fundamental nonlinear quantum theory additionally specifies mixed states, entangled systems, measurement, and production and detection. Deterministic nonlinear pure-state dynamics is subject to well-known signaling concerns under conventional ensemble and measurement assumptions~\cite{Gisin,Polchinski}. The causal framework of Ref.~\cite{Kaplan} illustrates the role of a complete construction; establishing the corresponding consistency properties for Eq.~\eqref{eq:action} requires its own ensemble and measurement framework. We consequently interpret the present result as a pure-state propagation benchmark with standard flavor preparation and detection assumed at leading order.

The chosen frame and stationarity are likewise substantive assumptions. Timelike fixed backgrounds select a preferred frame; directional backgrounds can introduce additional structures. The spacelike-$B$ example of Ref.~\cite{NgInfo} and rapid wave-packet gradients define extensions beyond the stationary identity in Eq.~\eqref{eq:secondzero}. The neutral-current contribution retained here is the leading Standard Model matter term; more precise matching could include radiative corrections. Antineutrino propagation follows by charge conjugating the microscopic nonlinear equation with a specified transformation rule for its external backgrounds. This procedure determines the nonlinear terms together with the familiar linear replacements $U\to U^*$ and $V\to-V$. These extensions build on the stationary neutrino results established here.

\section{Conclusion}
The flavor-singlet information-theoretic Dirac action leads, on its perturbative ultrarelativistic branch, to a normalized flavor equation with the restricted correction $-2g\langle\widehat C\rangle C$, where $C=2\Hv+\Hm$. The normalization follows from the action's conserved current and retains the derivative contribution at the order required by matching. A common mass-squared shift and flavor-universal neutral-current refraction enter the traceful coefficient even after all common propagation phases are removed.

The constant-density dynamics has a conserved functional and a commutator criterion for the evolution of its nonlinear strength. Its two-flavor resonance response is finite, and its second scalar-derivative sector vanishes exactly under the stationary, timelike-background assumptions. Reproducible numerical calculations check the effective-model conservation laws and perturbative response, and quantify the distinction between self-consistent, first-order, and frozen-coefficient evolution. Together with the analytic stationary-sector result, they turn the action into a definite propagation target. The correlated dependence on energy, source flavor, and matter composition guides phenomenological tests, while the identified degeneracies, scale dependence, and ensemble and measurement requirements define the route to a fundamental interpretation of their bounds.

\appendix
\section{Variation and the phase current}\label{app:variation}
Writing $\mathcal L_1=K^2/(4J)$, the relevant derivatives are
\begin{equation}
\frac{\partial\mathcal L_1}{\partial\bar N}
=\frac{K}{2J}i\gamma^\mu\partial_\mu N-\frac{K^2}{4J^2}A_\mu\gamma^\mu N,
\qquad
\frac{\partial\mathcal L_1}{\partial(\partial_\mu\bar N)}
=-i\frac{K}{2J}\gamma^\mu N.
\end{equation}
The Euler--Lagrange combination is precisely Eq.~\eqref{eq:n1}; in particular both copies of the derivative-on-$N$ term must be retained. For $\mathcal L_2=S^2/J$,
\begin{equation}
\frac{\partial\mathcal L_2}{\partial\bar N}
=2\xi B^\mu\partial_\mu N-\xi^2 A_\mu\gamma^\mu N,
\qquad
\frac{\partial\mathcal L_2}{\partial(\partial_\mu\bar N)}=2\xi B^\mu N,
\end{equation}
whose difference gives Eq.~\eqref{eq:n2} for constant $B$. Under the infinitesimal local phase change $N\to e^{i\alpha(x)}N$, $\delta J=\delta S=0$ and $\delta K=-2\bar N\gamma^\mu N\,\partial_\mu\alpha$. Thus $\delta\mathcal L=-f\bar N\gamma^\mu N\,\partial_\mu\alpha$, proving Eq.~\eqref{eq:current} on shell. This argument also establishes the role of the derivative term in Eq.~\eqref{eq:n1} when $\chi$ varies.

\bibliographystyle{unsrtnat}
\bibliography{references}

\begin{thebibliography}{25}
\providecommand{\natexlab}[1]{#1}
\providecommand{\url}[1]{\texttt{#1}}
\expandafter\ifx\csname urlstyle\endcsname\relax
  \providecommand{\doi}[1]{doi: #1}\else
  \providecommand{\doi}{doi: \begingroup \urlstyle{rm}\Url}\fi

\bibitem[Wolfenstein(1978)]{Wolfenstein}
L.~Wolfenstein.
\newblock Neutrino oscillations in matter.
\newblock \emph{Phys. Rev. D}, 17:\penalty0 2369--2374, 1978.
\newblock \doi{10.1103/PhysRevD.17.2369}.

\bibitem[Mikheyev and Smirnov(1985)]{MSW}
S.~P. Mikheyev and A.~Yu. Smirnov.
\newblock Resonance amplification of oscillations in matter and spectroscopy of
  solar neutrinos.
\newblock \emph{Sov. J. Nucl. Phys.}, 42:\penalty0 913--917, 1985.
\newblock Yad. Fiz. 42, 1441--1448 (1985).

\bibitem[Ng and Parwani(2009)]{NgDirac}
Wei~Khim Ng and Rajesh~R. Parwani.
\newblock Nonlinear {Dirac} equations.
\newblock \emph{SIGMA}, 5:\penalty0 023, 2009.
\newblock \doi{10.3842/SIGMA.2009.023}.

\bibitem[Ng and Parwani(2010)]{NgProbe}
Wei~Khim Ng and Rajesh~R. Parwani.
\newblock Probing quantum nonlinearities through neutrino oscillations.
\newblock \emph{Mod. Phys. Lett. A}, 25:\penalty0 793--804, 2010.
\newblock \doi{10.1142/S0217732310032792}.
\newblock URL \url{https://arxiv.org/abs/0805.3015}.

\bibitem[Ng and Parwani(2011)]{NgInfo}
Wei~Khim Ng and Rajesh~R. Parwani.
\newblock Information and particle physics.
\newblock \emph{Mod. Phys. Lett. A}, 26:\penalty0 681--689, 2011.
\newblock \doi{10.1142/S0217732311035146}.
\newblock URL \url{https://arxiv.org/abs/0908.0180}.

\bibitem[Cheung and Ng(2019)]{Cheung}
James~Asikin Cheung and Wei~Khim Ng.
\newblock Constraining non-linear {Dirac} equations with neutrino oscillations.
\newblock \emph{EPJ Web Conf.}, 206:\penalty0 09010, 2019.
\newblock \doi{10.1051/epjconf/201920609010}.

\bibitem[Quek et~al.(2020)Quek, Ng, Chan, and Oh]{Quek}
Zhi~Hao Quek, Wei~Khim Ng, Aik~Hui Chan, and Choo~Hiap Oh.
\newblock Nonlinear {Dirac} neutrino oscillations.
\newblock \emph{EPJ Web Conf.}, 240:\penalty0 07010, 2020.
\newblock \doi{10.1051/epjconf/202024007010}.

\bibitem[Kaplan and Rajendran(2022)]{Kaplan}
David~E. Kaplan and Surjeet Rajendran.
\newblock Causal framework for nonlinear quantum mechanics.
\newblock \emph{Phys. Rev. D}, 105:\penalty0 055002, 2022.
\newblock \doi{10.1103/PhysRevD.105.055002}.

\bibitem[Gherghetta and Shkerin(2023)]{Gherghetta}
Tony Gherghetta and Andrey Shkerin.
\newblock Out of this world neutrino oscillations.
\newblock \emph{Phys. Rev. D}, 108:\penalty0 075018, 2023.
\newblock \doi{10.1103/PhysRevD.108.075018}.

\bibitem[Duan et~al.(2006)Duan, Fuller, Carlson, and Qian]{Duan}
Huaiyu Duan, George~M. Fuller, J.~Carlson, and Yong-Zhong Qian.
\newblock Coherent development of neutrino flavor in the supernova environment.
\newblock \emph{Phys. Rev. Lett.}, 97:\penalty0 241101, 2006.
\newblock \doi{10.1103/PhysRevLett.97.241101}.

\bibitem[Lu(2026{\natexlab{a}})]{LuMatterFlows}
Jianlong Lu.
\newblock Invariant algebras of low-rank neutrino matter flows:
  {Krylov--Pl\"ucker} invariants, the {$3+1$} toric ring, and controlled
  deformations, 2026{\natexlab{a}}.
\newblock URL \url{https://arxiv.org/abs/2609.13915}.

\bibitem[Giunti(2007)]{Giunti}
Carlo Giunti.
\newblock Neutrino flavour states and the quantum theory of neutrino
  oscillations.
\newblock \emph{J. Phys. G}, 34:\penalty0 R93--R109, 2007.
\newblock \doi{10.1088/0954-3899/34/2/R02}.

\bibitem[Akhmedov and Wilhelm(2013)]{AkhmedovWilhelm}
Evgeny~Kh. Akhmedov and Alina Wilhelm.
\newblock Quantum field theoretic approach to neutrino oscillations in matter.
\newblock \emph{JHEP}, 01:\penalty0 165, 2013.
\newblock \doi{10.1007/JHEP01(2013)165}.

\bibitem[Dvornikov(2026)]{DvornikovDirac}
Maxim Dvornikov.
\newblock Quantum field theory treatment of oscillations of {Dirac} neutrinos
  in external fields.
\newblock \emph{Annals Phys.}, 487:\penalty0 170361, 2026.
\newblock \doi{10.1016/j.aop.2026.170361}.

\bibitem[Dvornikov(2025)]{Dvornikov}
Maxim Dvornikov.
\newblock Quantum field theory treatment of neutrino flavor oscillations in
  matter.
\newblock \emph{Phys. Rev. D}, 111:\penalty0 056009, 2025.
\newblock \doi{10.1103/PhysRevD.111.056009}.

\bibitem[Lu(2021)]{LuPhase}
Jianlong Lu.
\newblock Comment on: ``geometric phase of neutrinos: Differences between
  {Dirac} and {Majorana} neutrinos'' [{Phys. Lett. B} 780 (2018) 216].
\newblock \emph{Phys. Lett. B}, 818:\penalty0 136376, 2021.
\newblock \doi{10.1016/j.physletb.2021.136376}.

\bibitem[Esteban et~al.(2024)Esteban, Gonzalez-Garcia, Maltoni, Martinez-Soler,
  Pinheiro, and Schwetz]{NuFit}
Ivan Esteban, M.~C. Gonzalez-Garcia, Michele Maltoni, Ivan Martinez-Soler,
  Joao~Paulo Pinheiro, and Thomas Schwetz.
\newblock {NuFit-6.0}: Updated global analysis of three-flavor neutrino
  oscillations.
\newblock \emph{JHEP}, 12:\penalty0 216, 2024.
\newblock URL \url{https://arxiv.org/abs/2410.05380}.

\bibitem[Ohlsson(2013)]{Ohlsson}
Tommy Ohlsson.
\newblock Status of non-standard neutrino interactions.
\newblock \emph{Rep. Prog. Phys.}, 76:\penalty0 044201, 2013.
\newblock \doi{10.1088/0034-4885/76/4/044201}.

\bibitem[Aker et~al.(2025)]{KATRIN259}
M.~Aker et~al.
\newblock Direct neutrino-mass measurement based on 259 days of {KATRIN} data.
\newblock \emph{Science}, 388:\penalty0 180--185, 2025.
\newblock \doi{10.1126/science.adq9592}.

\bibitem[Abi et~al.(2020)]{DUNEPhysics}
B.~Abi et~al.
\newblock Long-baseline neutrino oscillation physics potential of the {DUNE}
  experiment.
\newblock \emph{Eur. Phys. J. C}, 80:\penalty0 978, 2020.
\newblock \doi{10.1140/epjc/s10052-020-08456-z}.

\bibitem[Agostini et~al.(2018)]{BorexinoPP}
M.~Agostini et~al.
\newblock Comprehensive measurement of pp-chain solar neutrinos.
\newblock \emph{Nature}, 562:\penalty0 505--510, 2018.
\newblock \doi{10.1038/s41586-018-0624-y}.

\bibitem[Abusleme et~al.(2026)]{JUNOFirst}
A.~Abusleme et~al.
\newblock Measurement of reactor neutrino oscillation with the first {JUNO}
  data.
\newblock \emph{Nature}, 654:\penalty0 343--348, 2026.
\newblock \doi{10.1038/s41586-026-10538-z}.

\bibitem[Lu(2026{\natexlab{b}})]{LuEstimation}
Jianlong Lu.
\newblock Operational quantum estimation theory for neutrino oscillations:
  identifiability, attainability, and spectral precision bounds,
  2026{\natexlab{b}}.
\newblock URL \url{https://arxiv.org/abs/2608.23654}.

\bibitem[Gisin(1990)]{Gisin}
Nicolas Gisin.
\newblock Weinberg's non-linear quantum mechanics and supraluminal
  communications.
\newblock \emph{Phys. Lett. A}, 143:\penalty0 1--2, 1990.
\newblock \doi{10.1016/0375-9601(90)90786-N}.

\bibitem[Polchinski(1991)]{Polchinski}
Joseph Polchinski.
\newblock Weinberg's nonlinear quantum mechanics and the
  {Einstein-Podolsky-Rosen} paradox.
\newblock \emph{Phys. Rev. Lett.}, 66:\penalty0 397--400, 1991.
\newblock \doi{10.1103/PhysRevLett.66.397}.

\end{thebibliography}
\end{document}